

\magnification=\magstep1

\vsize=9.0truein
\hsize=6.1truein
\vskip 0.25in
\baselineskip=14pt plus 0.2pt minus 0.3pt
\parskip=12pt
\parindent=0pt
\hoffset=0.3in
\footline={\ifnum\pageno<2\hfil\else\hss\tenrm\folio\hss\fi}
\font\bigrm=cmr10 scaled\magstep3
\def\subE{{\scriptscriptstyle E}}
\def\H{{\cal H}}
\def\D{{\cal D}}
{\bigrm Jacobi's Action and the Density of States}
\vskip 5pt
{\bigrm J. David Brown}\hfil\break
Departments of Physics and Mathematics\hfil\break
North Carolina State University, Raleigh, NC 27695--8202
\vskip 5pt
{\bigrm James W. York, Jr.}\hfil\break
Department of Physics and Astronomy\hfil\break
University of North Carolina, Chapel Hill, NC 27599--3255

The authors have introduced recently a ``microcanonical functional
integral" which yields directly the density of states as a function
of energy. The phase of the functional integral is Jacobi's action,
the extrema of which are classical solutions at a given energy. This
approach is general but is especially well suited to gravitating
systems because for them the total energy can be fixed simply as a
boundary condition on the gravitational field. In this paper, however,
we ignore gravity and illustrate the use of Jacobi's action by
computing the density of states for a nonrelativistic harmonic
oscillator.

{\bf 1 DEDICATION}\hfil\break
We dedicate this paper to Dieter Brill in honor of his sixtieth
birthday. His continued fruitful research in physics and his personal
kindness make him a model colleague. JWY would especially like to
thank him for countless instructive discussions and for his friendship
over the past twenty--five years.

{\bf 2 INTRODUCTION}\hfil\break
Jacobi's form of the action principle involves variations at fixed
energy, rather than the variations at fixed time used in Hamilton's
principle. The fixed time interval in Hamilton's action becomes fixed
inverse temperature in the ``periodic imaginary time" formulation,
thus transforming Hamilton's action into the appropriate (imaginary)
phase for a periodic path in computing the canonical partition
function from a Feynman functional integral (Feynman and Hibbs 1965).
In contrast, fixed total energy is suitable for the microcanonical
ensemble and, correspondingly, Jacobi's action is the phase in an
expression for the density of states as a real--time ``microcanonical
functional integral" (MCFI) (Brown and York 1993b).

We wish to characterize briefly the canonical and microcanonical
pictures. (We shall speak only of energy and (inverse) temperature
here, ignoring the other possible conjugate pairs of variables in
order to simplify the discussion.) In the canonical picture, with a
fixed temperature shared
by all constituents of a system, there are no constraints on the
energy. This feature simplifies combinatorial (counting) problems for
canonical systems and leads to the factorization of the partition
function for weakly coupled constituents. For gravitating systems in
equilibrium, the temperature is not spatially uniform because
of gravitational red and blue--shift effects. In such cases the
relevant temperature is that determined at the boundary of the system
(York 1986). It can therefore be specified as a boundary condition on
the metric (York 1986, Whiting and York 1988, Braden {\it et
al\/}.~1990) and used in conjunction with Hamilton's principle, which
is the form of the action for gravity in which the metric is fixed
on the boundary (Brown and York 1992, 1993a). (The
metric determines the lapse of proper time along the boundary.) On
the other hand, equilibrium in the canonical picture is not always
stable when gravity is present, as is well known. For some pertinent
examples, see York (1986), Whiting and York (1988), and Braden {\it
et al\/}.~(1990).

With its constraint on the energy, the microcanonical picture leads
to more robust stability properties.  However, the energy constraint
can complicate calculations of relevant statistical properties
because the constituents of the system share from a common fixed pool
of energy. For field theories, with a continuous infinity of degrees
of freedom, the energy constraint restricts  the entire phase space
of the system {\it unless gravity is taken into account\/}. For
gravitating systems, as a  consequence of the equivalence principle,
the total energy including that of matter fields is an integral of
certain derivatives of the metric over a two--surface bounding the
system. Therefore, if we specify as a boundary condition the energy
per unit two--surface area, we have constrained the total energy
simply by a boundary condition (Brown and York 1993a, 1993b).
Thus, the canonical and microcanonical cases differ only in which of
the conjugate variables (Brown {\it et al\/}.~1990), inverse
temperature or energy, is specified on the boundary. The corresponding
functional integrals, for partition function or density of states,
differ in which action gives the correct phase, Hamilton's or Jacobi's.

We have recently applied this reasoning to the case of a stationary
black hole (Brown {\it et al\/}.~1991a, 1991b, Brown and York 1993b).
The MCFI, in a steepest descents approximation, shows that the
logarithm of the density of states is one--quarter of the area
of the event horizon (that is, the Bekenstein--Hawking entropy) (Brown
and York 1993b). In the present paper we shall disregard gravity and
obtain the density of states for a nonrelativistic harmonic
oscillator. This is a relatively simple situation in which to recall
the properties of Jacobi's action and to see the MCFI at work.

\vfill\eject
{\bf 3 JACOBI'S ACTION}\hfil\break
Consider, for simplicity, a particle of mass $m$ with a
one--dimensional configuration space. The Lagrangian form of Jacobi's
action is (Lanczos 1970, Brown and York 1989)
$$S_{\subE}[x] = \int dx \sqrt{2m\bigl[ E-V(x) \bigr]} \ ,\eqno(1)$$
where $V(x)$ is the potential energy and the energy $E$ is a fixed
constant. $S_\subE[x]$ is extremized by varying the path freely except
that the end points are fixed. Now introduce a parameter $\sigma$
increasing monotonically from $\sigma'$ at one end of the path to
$\sigma''$ at the other. Denoting $dx/d\sigma$ by $\dot x$, we can
write the action as
$$S_{\subE}[x] = \int_{\sigma'}^{\sigma''} d\sigma \,\dot x
      \sqrt{2m\bigl [ E-V(x) \bigr]} \ ,\eqno(2)$$
where $x' = x(\sigma')$ and $x''= x(\sigma'')$ are fixed. Jacobi's
action is invariant under changes $\delta x$ induced by changes of
parameterization that preserve the end--point values of $\sigma$.

For constructing the MCFI, we employ the canonical form of Jacobi's
action. Because of the reparameterization invariance of $S_\subE[x]$,
the corresponding canonical Hamiltonian $\dot x(\partial L/\partial\dot
x) - L$ vanishes identically. Furthermore, the canonical momentum
$$p = {\partial L\over\partial\dot x} = \bigl[2m(E-V)\bigr]^{1/2}
   \eqno(3)$$
is independent of $\dot x$ in one dimension and, in general, does not
allow one to solve for all the $\dot x$'s as functions of the $p$'s.
Indeed, from (3) we obtain the ``Hamiltonian constraint"
$$\H(x,p) \equiv {p^2\over 2m} + V(x) - E \approx 0 \ .\eqno(4)$$
Because the canonical Hamiltonian is zero, there are no secondary
constraints and $\H$ is then trivially first class. Jacobi's action
in canonical form is thus
$$S_\subE[x,p,N] = \int_{\sigma'}^{\sigma''} d\sigma \bigl[ p\dot x
       - N\H(x,p)\bigr] \ ,\eqno(5)$$
where $N$ is a Lagrange multiplier.
The equations of motion following from variation of (5) are
$$\eqalignno{ \dot x &= N[x,\H] = {Np\over m} &(6)\cr
        \dot p &= N[p,\H] = -N {\partial V\over\partial x} &(7)\cr
        \H &= {p^2\over 2m} +V -E =0 \ .&(8)\cr}$$
Combining (6) and (8) determines the multiplier as
$$N = \dot x \bigl[ 2(E-V)/m\bigr]^{-1/2} \ .\eqno(9)$$
The interpretation of (9) is that
$$dt = N\, d\sigma \eqno(10)$$
is the lapse of physical time, in accordance with the definition
of energy.

The canonical statement of reparameterization invariance is that the
action (5) is invariant under the gauge transformation given by
$$\eqalignno{ \delta x &= \epsilon [x,\H] \ ,&(11)\cr
              \delta p &= \epsilon [p,\H] \ ,&(12)\cr
              \delta N &= \dot\epsilon \ ,&(13)\cr}$$
where $\epsilon(\sigma') = \epsilon(\sigma'') =0$. With the choice
$$\epsilon(\sigma) = \biggl({\sigma-\sigma'\over\sigma''
     -\sigma'}\biggr)
    T - \int_{\sigma'}^{\sigma} d\alpha\, N(\alpha) \ ,\eqno(14)$$
where $T$ is the total time
$$T= \int_{\sigma'}^{\sigma''} d\sigma\, N(\sigma) \ ,\eqno(15)$$
the lapse function is transformed to a constant, namely, $N =
T/(\sigma''- \sigma')$. This shows that every history is gauge
related to a history with a constant lapse, and the time $T$ is the
gauge invariant part of the lapse function. If the histories under
consideration are restricted to those with constant lapse, the gauge
freedom of Jacobi's action is removed and (5) becomes
$$S_\subE[x,p;T) = \int_{\sigma'}^{\sigma''} d\sigma \bigl[ p\dot x
   - T  \H /(\sigma''-\sigma') \bigr] \ .\eqno(16)$$
This form of Jacobi's action is a functional of $x(\sigma)$ and
$p(\sigma)$ and an ordinary function of the time interval $T$. The
classical equations of motion for (16), that is, the conditions for
the extrema of (16), are given by (6) and (7) with $N=T/(\sigma''-
\sigma')$ along with
$$0 = {\partial S_\subE \over \partial T}  = - {1\over(\sigma''
    -\sigma')} \int_{\sigma'}^{\sigma''}   d\sigma\,\H \ .\eqno(17)$$
Since (6) and (7) imply that $\H$ is constant, equations (6), (7), and
(17) together imply $\H=0$. It follows that the form (16) for Jacobi's
action is classically equivalent to (5), but has no gauge freedom.

{\bf 4 FUNCTIONAL INTEGRAL FOR JACOBI'S ACTION}\hfil\break
The functional integral associated with Jacobi's action can be
constructed by integrating over all histories $x(\sigma)$, $p(\sigma)$,
$T$, with fixed endpoints $x(\sigma')=x'$ and $x(\sigma'')=x''$, where
the phase for each history is given by the action (16). Thus, the
functional integral is
$$Z_\subE(x'',x') = {1\over2\pi\hbar} \int dT
     \int_{x(\sigma')=x'}^{x(\sigma'')=x''} \D x\D p \exp\biggl\{
     {i\over\hbar} \int_{\sigma'}^{\sigma''} d\sigma \bigl[ p\dot x -
    T\H/(\sigma''-\sigma')  \bigr] \biggr\} \ ,\eqno(18)$$
where $\D x\D p$ is (formally) the product over $\sigma$ of the
Liouville phase space measure $dx(\sigma)dp(\sigma)/(2\pi\hbar)$. The
integration measure in (18) can be justified by appealing to a BRST
analysis based on the canonical action (5), as is done in the Appendix
of Brown and York (1993b).

The functional integral over $x(\sigma)$ and $p(\sigma)$ in (18)
has the familiar form of the path integral associated with Hamilton's
action, where $\sigma$ plays the role of time and the Hamiltonian is
$T\H/(\sigma''-\sigma')$. This path integral can be written as the
matrix elements of the evolution operator $\exp(-iT\hat\H/\hbar)$, so
the path integral for Jacobi's action becomes
$$Z_\subE(x'',x') = {1\over2\pi\hbar} \int dT <x''|e^{-iT\hat\H/\hbar}
    |x'>  \ .\eqno(19)$$
Hence, taking the integration of $T$ over all real values, we have
$$Z_\subE(x'',x') = <x''|\delta(\hat\H)|x'> \ .\eqno(20)$$
Note that $Z_\subE(x'',x')$ satisfies the time independent
Schr\"odinger equation, namely $\hat\H Z_\subE(x'',x') = 0$ (where
$\hat\H$ acts on the argument $x''$), since formally
$\hat\H\delta(\hat\H) = 0$.

{}From (20) it follows that the trace of $Z_\subE(x'',x')$ yields
the density of states
$$\nu(E) = \int dx\, Z_\subE(x,x) = {\rm Tr}\delta(\hat\H)
   = {\rm Tr}\delta(E-\hat H) \ ,\eqno(21)$$
where $\hat H$ is the usual Hamiltonian operator.
By combining this result with (18) we find that $\nu(E)$ can be
written directly as a functional integral, the MCFI:
$$\nu(E) = {1\over2\pi\hbar} \int dT \int \D x\D p \exp\biggl\{
     {i\over\hbar} \int_{-\pi}^{\pi} d\sigma \bigl[ p\dot x -
    T\H/2\pi \bigr] \biggr\} \ .\eqno(22)$$
For later convenience, the endpoint parameter values have been chosen
to be $\sigma' = -\pi$ and $\sigma'' = \pi$. The derivation of the
path integral (22) for the density of states shows that the
integration can be described as a sum over all phase space curves that
begin and end at some ``base point" $x(\pi) = x(-\pi) = x$, plus an
integral over the base point $x$. Then roughly speaking, the density
of states is given by a sum over all periodic histories. However, to
be precise, it should be recognized that the sum in (22) counts each
closed phase space curve a continuous infinity of times because any
point on the curve can serve as the base point $x$. Also observe that
the integration in (22) is over all real values of the time interval
$T$, rather than just positive values. This implies that the
functional integral for $\nu(E)$ consists of a sum over {\it pairs\/}
of histories with members contributing equal and opposite phases
(Brown and York 1993b). As a consequence, the density of
states so constructed is real.

{\bf 5 DENSITY OF STATES FOR THE HARMONIC OSCILLATOR}\hfil\break
We now turn to the evaluation of the MCFI (22) for
the density of states of a simple harmonic oscillator with angular
frequency $\omega$ and Hamiltonian constraint
$$\H = {p^2\over 2m} + {m\omega^2 x^2\over 2} -E \ .\eqno(23)$$
The periodic nature of the histories suggests the use of Fourier
series techniques (Feynman and Hibbs 1965) for this calculation.
Accordingly, write the phase space coordinates as
$$\eqalignno{ x(\sigma) &= a_0 + \sum^\infty_{k=1}\bigl( a_k\cos
    k\sigma  + b_k\sin k\sigma\bigr) \ ,&(24)\cr
       p(\sigma) &= c_0 + \sum^\infty_{k=1}\bigl( c_k\cos k\sigma
   + d_k\sin k\sigma\bigr) \ .&(25)\cr}$$
The functional integral over $x(\sigma)$ and $p(\sigma)$ is replaced
by a multiple integral over the coefficients in the Fourier series
(24) and (25) with measure
$$\D x\D p = J\,da_0\,dc_0\,\prod_{k=1}^\infty \bigl(
      da_k\,db_k\,dc_k\,dd_k \bigr) \ .\eqno(26)$$
Here, $J$ is (formally) the Jacobian of the transformation from
$x(\sigma)$, $p(\sigma)$ to $a_0$, $c_0$, $a_k$, $b_k$, $c_k$, $d_k$.
The form (24), (25) of this transformation shows (again, formally)
that $J$ should be a {\it real constant\/}, and should be independent
of $T$, $m$, $\omega$, and $E$. ($J$ should depend on $\hbar$, since
$\hbar$ appears in the definition of $\D x\D p$.) One of the goals of
the present calculation is to determine the real constant $J$ that
characterizes the change of integration variables specified by (24)
and (25). Note that by integrating freely over all Fourier coefficients
$a_0$, $c_0$, $a_k$, $b_k$, $c_k$, $d_k$, we have each closed phase
space curve correctly counted a continuous infinity of times. This is
because the values of the Fourier coefficients depend on the choice of
base point that is assigned the parameter value $\sigma=\pi$
(identified with $\sigma=-\pi$) on a given closed phase space curve.

With the change of variables (24), (25), the density of states (22)
for the harmonic oscillator becomes
$$\nu(E) = {J\over2\pi\hbar} \int dT \int da_0\,dc_0\,
   \prod_{k=1}^\infty  \bigl( da_k\,db_k\,dc_k\,dd_k \bigr)
   \exp\bigl\{iS_\subE/\hbar\bigr\}   \ ,\eqno(27)$$
where the phase is obtained by substituting the Fourier series for
$x(\sigma)$ and $p(\sigma)$ into the action (16):
$$\eqalignno{ S_\subE &= ET - {m\omega^2 T\over2} a_0^2 - {T\over2m}
      c_0^2 \cr
      &\ -{1\over2}\sum_{k=1}^\infty \biggl\{ 2\pi k(a_k d_k -
    b_k c_k) +  {T\over2m}(c_k^2 + d_k^2) + {m\omega^2 T\over2}(a_k^2
    + b_k^2)    \biggr\} \ .&(28)\cr}$$
The calculation is simplified by expanding $c_k$ and $d_k$ about the
solutions to their ``equations of motion". Accordingly, observe that
the action (28) is extremized for $c_k$ and $d_k$ that satisfy
$$\eqalignno{ 0 &= {\partial S_\subE\over\partial c_k} = \pi k b_k
            -{T\over2m} c_k \ ,&(29)\cr
              0 &= {\partial S_\subE\over\partial d_k} = -\pi k a_k
            -{T\over2m} d_k \ .&(30)\cr}$$
Thus, define new integration variables $\bar c_k$ and $\bar d_k$ by
$$\eqalignno{ c_k &= {2\pi m\over T}kb_k + \bar c_k \ ,&(31)\cr
           d_k &= -{2\pi m\over T}ka_k + \bar d_k \ ,&(32)\cr}$$ and
the action (28) becomes
$$\eqalignno{ S_\subE &= ET - {m\omega^2 T\over2} a_0^2 - {T\over2m}
       c_0^2 \cr
      &\ -{1\over2}\sum_{k=1}^\infty \biggl\{ {T\over2m}(\bar c_k^2 +
   \bar d_k^2)   + {m\over2T}(\omega^2T^2 - 4\pi^2k^2)(a_k^2 + b_k^2)
   \biggr\}   \ .&(33)\cr}$$
The integrations over $a_0$, $c_0$, $a_k$, $b_k$, $\bar c_k$, and
$\bar d_k$ are now straightforward since these variables are uncoupled
in the action (33). Moreover, for each value of $k$, the integrals over
$\bar c_k$ are identical to the integrals over $\bar d_k$, and the
integrals over $a_k$ are identical to the integrals over $b_k$. From
these observations it follows
that the density of states (27) can be written as
$$\eqalignno{ \nu(E) &= {J\over2\pi\hbar} \int dT\,da_0dc_0
      \exp\biggl\{
      {i\over\hbar}\biggl[ ET - {m\omega^2 T\over2} a_0^2 -
      {T\over2m} c_0^2 \biggr]\biggr\} \cr
      &\ \times\biggl( \prod_{k=1}^\infty \int da_k d\bar c_k
       \exp\biggl\{ {i\over\hbar}\biggl[
       {m(4\pi^2k^2 - \omega^2 T^2 )\over4T} a_k^2 -
        {T\over4m} \bar c_k^2 \biggr]\biggr\}\biggr)^2
       \ .&(34)\cr}$$
Each of these integrals (excluding the integral over $T$) has the form
of a Fresnel integral,
$$\int dx \exp\bigl(iAx^2\bigr) = \sqrt{{\pi\over |A|}} \exp\bigl(
     i\pi{\rm sign}A/4  \bigr) \ ,\eqno(35)$$
where the constant $A$ is real. In evaluating (34), it is helpful to note
that the square of the Fresnel integral (35) is $i\pi/A$. The result is
$$\nu(E) = -iJ\int dT {1\over\omega T}  \prod_{k=1}^\infty
           \biggl[ \biggl({2\hbar\over k}\biggr)^2 \biggl( 1 - {\omega^2
    T^2  \over 4\pi^2 k^2} \biggr)^{-1}\biggr]
        \exp\biggl\{ {i\over\hbar} ET\biggr\} \ .\eqno(36)$$
Now use the identity
$$\sin x = x\prod_{k=1}^\infty \biggl(1-{x^2\over\pi^2k^2}\biggr)
\eqno(37)$$
to obtain
$$\nu(E) = {-iJ\over2} \prod_{k=1}^\infty \biggl({2\hbar\over k}
    \biggr)^2  \int dT {1\over \sin(\omega T/2)} \exp\biggl\{
    {i\over\hbar} ET\biggr\}   \ .\eqno(38)$$
Next, express the inverse of $\sin(\omega T/2)$ as
$$\eqalignno{ {1\over \sin(\omega T/2)} &= {2i\over e^{i\omega T/2} -
  e^{-i\omega T/2}} = 2i e^{-i\omega T/2} {1\over 1-e^{-i\omega T}} \cr
    &= 2i e^{-i\omega T/2} \sum_{n=0}^\infty e^{-i\omega Tn}
   \ ,&(39)\cr}$$
and insert this result into (38). Integrating the series
term--by--term, we obtain
$$\nu(E) = 2\pi\hbar J \prod_{k=1}^\infty \biggl({2\hbar\over k}
   \biggr)^2  \sum_{n=0}^\infty \delta\bigl( E - \hbar\omega(n+1/2)
   \bigr) \ .\eqno(40)$$
This result shows that the Jacobian $J$ for the change of variables
(24), (25) should be identified with the (real, infinite) constant
$$J = {1\over2\pi\hbar} \prod_{k=1}^\infty \biggl({k\over2\hbar}
   \biggr)^2  \ .\eqno(41)$$
Then the density of states becomes
$$\nu(E) = \sum_{n=0}^\infty \delta\bigl( E - \hbar\omega(n+1/2)
   \bigr) \ ,\eqno(42)$$
which is the anticipated result showing that for the harmonic
oscillator $\nu(E)$ is a sum of delta functions peaked at
half--odd--integer multiples of $\hbar\omega$.

Finally, we note that the various quantum--statistical and
thermodynamical properties of a system can be obtained from its
density of states. In particular, the canonical partition function
$Z(\beta)$ is defined as the Laplace transform of $\nu(E)$, and from
$Z(\beta)$ the heat capacity, entropy, and other thermodynamical
quantities can be found. For the harmonic oscillator with density
of states (42), the partition function is
$$Z(\beta) = \int_0^\infty dE\,\nu(E) \, e^{-\beta E} =
    \sum_{n=0}^\infty  e^{-\beta\omega\hbar(n+1/2)} \ ,\eqno(43)$$
which is the well known result.

{\bf ACKNOWLEDGMENTS}\hfil\break
 This research was supported by National Science Foundation grant
PHY--8908741.

{\bf REFERENCES}

\hangindent=10pt\hangafter=1
Braden, H. W., J. D. Brown, B. F. Whiting, and J. W. York (1990).
Charged black hole in a grand canonical ensemble. {\it Physical Review\/},
{\bf D42}, 3376--3385.

\hangindent=10pt\hangafter=1
Brown, J. D. and J. W. York (1989). Jacobi's action and the recovery
of time in general relativity. {\it Physical Review\/}, {\bf D40},
3312--3318.

\hangindent=10pt\hangafter=1
Brown, J. D., G. L. Comer, E. A. Martinez, J. Melmed, B. F.
Whiting, and J. W. York (1990). Thermodynamic ensembles and gravitation. {\it
Classical and Quantum Gravity}, {\bf 7}, 1433--1444.

\hangindent=10pt\hangafter=1
Brown, J. D., E. A. Martinez, and J. W. York (1991a). Rotating black holes,
complex geometry, and thermodynamics. In {\it Nonlinear Problems in Relativity
and Cosmology\/}, Eds J. R. Buchler, S. L. Detweiler, and J. R. Ipser. New
York Academy of Sciences, New York.

\hangindent=10pt\hangafter=1
Brown, J. D., E. A. Martinez, and J. W. York (1991b). Complex Kerr--Newman
geometry and black hole thermodynamics. {\it Physical Review Letters\/},
{\bf 66}, 2281--2284.

\hangindent=10pt\hangafter=1
Brown, J. D. and J. W. York (1992). Quasi-local energy in
general relativity. In {\it Mathematical Aspects of Classical Field Theory},
Eds
M. J. Gotay, J. E. Marsden, and V. E. Moncrief. American Mathematical Society,
Providence.

\hangindent=10pt\hangafter=1
Brown, J. D. and J. W. York (1993a). Quasilocal energy and conserved
charges derived from the gravitational action. To appear in {\it Physical
Review\/}, {\bf D47}.

\hangindent=10pt\hangafter=1
Brown, J. D. and J. W. York (1993b). Microcanonical functional integral for
the gravitational field. To appear in {\it Physical Review\/}, {\bf D47}.

\hangindent=10pt\hangafter=1
Feynman, R. P. and A. R. Hibbs (1965). {\it Quantum Mechanics and Path
Integrals\/}. McGraw Hill, New York.

\hangindent=10pt\hangafter=1
Lanczos, C. (1970). {\it The Variational Principles of Mechanics\/}.
University of Toronto Press, Toronto.

\hangindent=10pt\hangafter=1
Whiting, B. F. and J. W. York (1988). Action principle and partition
function for the gravitational field in black--hole topologies. {\it Physical
Review Letters\/}, {\bf 61}, 1336--1339.

\hangindent=10pt\hangafter=1
York, J. W. (1986). Black-hole thermodynamics and the Euclidean
Einstein action. {\it Physical Review}, {\bf D33}, 2092--2099.
\bye